\documentclass[apj]{emulateapj}

\usepackage{graphicx,times}
\usepackage{subfigure}
\newcommand{\be}{\begin{equation}}
\usepackage{threeparttable}
\usepackage{booktabs}
\newcommand{\ee}{\end{equation}}
\newcommand{\bea}{\begin{eqnarray}}
\newcommand{\eea}{\end{eqnarray}}

\usepackage{enumerate}
\usepackage{amsmath}
\usepackage{cases}
\usepackage{longtable}
\usepackage{hyperref}
\usepackage{epstopdf}
\usepackage{amsmath,bm}
\usepackage{amssymb}
\usepackage{natbib}
\usepackage{morefloats}
\usepackage{multirow}
\usepackage{array}
\usepackage{verbatim}

\begin{document}

\title{Trapping of cosmic rays in MHD turbulence}

\author{Siyao Xu\altaffilmark{1,2} and Alex Lazarian\altaffilmark{1} }

\altaffiltext{1}{Department of Astronomy, University of Wisconsin, 475 North Charter Street, Madison, WI 53706, USA; 
sxu93@wisc.edu,
lazarian@astro.wisc.edu}
\altaffiltext{2}{Hubble Fellow}

\begin{abstract}

Astrophysical plasmas are turbulent and magnetized. 
The interaction between cosmic rays (CRs) and 
magnetohydrodynamic (MHD) turbulence is a fundamental astrophysical process. 
Based on the current understanding of MHD turbulence, 
we revisit the trapping of CRs by magnetic mirrors in the context of MHD turbulence. 
In compressible MHD turbulence, isotropic fast modes dominate both trapping and gyroresonant scattering of CRs. 
The presence of trapping significantly suppresses the pitch-angle scattering and the 
spatial diffusion of CRs along the magnetic field. 
The resulting parallel diffusion coefficient has a weaker dependence on CR energy at higher energies. 
In incompressible MHD turbulence, 
the trapping by pseudo-Alfv\'{e}n modes dominates over the gyroresonant scattering by anisotropic Alfv\'{e}n and pseudo-Alfv\'{e}n modes
at all pitch angles
and prevents CRs from diffusion.

\end{abstract}

\keywords{turbulence - magnetic fields - cosmic rays }

\section{Introduction}

As important cosmic messengers,
cosmic rays (CRs) carry unique information about their sources and the media that they propagate through. 
The problem of the origin of CRs concerns their spatial diffusion
\citep{Kulsrud_Pearce,Am14},
which remains a great challenge for modern astrophysics. 
For the reconstruction of CR diffusion, a proper statistical description of the turbulent magnetic fields that CRs interact with
is crucial,
which requires both ever-improving observations and theoretical efforts.

Historically, theoretical studies on the scattering and diffusion of CRs were based on phenomenological models of turbulent magnetic fields
{(e.g., \citealt{Mat90,Giacalone_Jok1999,Sha16})}
and the quasi-linear theory (QLT) by assuming unperturbed orbits of particles 
\citep{Jokipii1966}.
Comparisons of the theoretical expectations with simulations 
(e.g. \citealt{Qin2002})
and observations collected over the last decade
show apparent discrepancies  
(see the review by \citealt{Gabi19}).
More recently, realistic models for magnetohydrodynamic (MHD) turbulence
have been established and numerically tested 
\citep{GS95, LV99, CV00,MG01,CLV_incomp, CL03, KL12, Bere14}, 
which are also supported by observations in the solar wind
\citep{Hor08,For11}.
The modern theories of MHD turbulence bring radical changes of the standard diffusive paradigm of CRs
\citep{Chan00,YL04,Brunetti_Laz,XY13, LY14,Xuc16,XLb18,Sio20}.

The application of the QLT to anisotropic MHD turbulence leads to inefficient scattering of CRs
\citep{Chan00,YL02}.
Different mechanisms, including the nonlinear resonance-broadened transit time damping (TTD) 
\citep{YL08,XLb18,Dem19}
and the streaming instability excited by low-energy CRs
\citep{Lerche,Kulsrud_Pearce}, 
have been invoked to enhance the scattering and 
confine the diffusion of CRs. 
Besides the scattering of CRs by MHD turbulence 
{(see the review by 
\citealt{Mer19}
for test particle simulations of CRs)}
and self-excited {turbulence/instabilities
\citep{Bla12,Leb18,Bai19,Hol19,Krum20},} 
trapping of CRs in compressible MHD turbulence in, e.g., the solar wind, the interstellar medium (ISM), and the intracluster medium,
can also significantly affect the diffusion of CRs. 
The magnetic compressions with the field variation scale larger than the CR gyroradius
act as magnetic mirrors, trapping the CRs that conserve their first adiabatic invariant. 
This trapping effect can also remove the singularity in parallel diffusion coefficient at $90^\circ$ 
\citep{CesK73},
which is a fundamental difficulty of the QLT
\citep{Jokipii1966}.
Trapping of CRs by large-scale magnetic irregularities was earlier studied by, e.g., 
{\citet{Fer49,Noe68,CesK73,Klep95,Zira01,Med15},} 
but it has not been investigated in the framework of modern theories of MHD turbulence.

In this work, we focus on the trapping of CRs in MHD turbulence and examine the scattering and diffusion of CRs in the 
presence of trapping. 
In Section 2, we analyze
the gyroresonant scattering of CRs by Alfv\'{e}n, slow, and fast modes of MHD turbulence.
In Section 3, we study the effect of trapping on CR diffusion. 
A discussion is presented in Section 4. 
Finally, the summary of our main results is given in Section 5.


\section{Pitch-angle scattering by MHD turbulence }
\label{sec: pasca}

Compressible MHD turbulence can be decomposed into Alfv\'{e}n, slow, and fast modes 
\citep{CL03}.
Alfv\'{e}n modes in compressible MHD turbulence 
have the same scale-dependent anisotropy as those in incompressible MHD turbulence 
\citep{GS95}
in the local frame of the magnetic field 
\citep{LV99,CV00,MG01}.
The anisotropic scaling also applies to slow modes and pseudo-Alfv\'{e}n modes in the incompressible limit, 
as they are passively mixed by the cascade of Alfv\'{e}n modes 
\citep{LG01}.
Fast modes have independent energy cascade and isotropic scaling
\citep{CL02_PRL}.

(1) Alfv\'{e}n modes. 
For describing the gyroresonant interactions with Alfv\'{e}n modes, 
the pitch-angle diffusion coefficient is 
\citep{Volk:1975},
\begin{equation}\label{eq: oriduav}
     D_{\mu\mu,A} = C_\mu \int d^3k x^{-2} [J_1(x)]^2 I_A(k) R(k),
\end{equation}
with
\begin{equation}\label{eq: cmu}
    C_\mu  = (1 - \mu^2) \frac{\Omega^2}{B_0^2}, 
\end{equation}
and 
\begin{equation} \label{eq: asux}
     x = \frac{k_\perp v_\perp}{ \Omega}  = \frac{k_\perp }{r_g^{-1} },
\end{equation}
where $\Omega$ is the gyrofrequency, $r_g = v_\perp / \Omega$ is the gyroradius, 
$v$ is the particle speed, $B_0$ is the strength of mean magnetic field, 
$\mu = v_\| / v$ is the pitch-angle cosine, 
and $\|$ and $\perp$ 
denote directions with respect to the local magnetic field. 
In addition,  in the quasilinear approximation
the resonance function for gyroresonance is 
\begin{equation}\label{eq: qltreg}
    R_L = \pi \delta(\omega_k - v_\| k_\| + \Omega )  , 
\end{equation}
where $\omega_k$ is the wave frequency, and it is negligible compared with $v_\| k_\|$ for relativistic particles. 
As a proper description of the scaling properties of MHD turbulence, we adopt the 
magnetic energy spectrum of Alfv\'{e}nic turbulence tested by 
\citet{CLV_incomp},
\begin{equation}\label{eq: enespa}
     I_A(k) = C_A  k_\perp^{-\frac{10}{3}} \exp{\Bigg(-L^\frac{1}{3}\frac{k_\|}{k_\perp^\frac{2}{3}}\Bigg)},
\end{equation} 
with the normalization factor 
\begin{equation}
    C_A = \frac{1}{6 \pi} \delta B_A^2 L^{-\frac{1}{3}},
\end{equation}
where $L$ is the injection scale of turbulence, and $\delta B_A$ is the rms strength of the fluctuating magnetic fields of Alfv\'{e}n modes at $L$. 
{The normalization factor used here and in the rest of the paper 
is chosen to have the integral of the magnetic energy spectrum over wavenumber space equal to 
$\delta B^2 / 2$, where $\delta B$ is the rms strength of the fluctuating magnetic fields of each modes at $L$.
\footnote{{Different normalizations can be adopted by different authors (see, e.g., \citealt{Schlickeiser02}).} }}
We note that in the case of super-Alfv\'{e}nic turbulence with the injected turbulent energy larger than the magnetic energy, 
$L$ in Eq. \eqref{eq: enespa} should be replaced by the Alfv\'{e}nic scale $l_A = L M_A^{-3}$, where $M_A = V_L/ V_A$ is the Alfv\'{e}n Mach number, $V_L$ is the injected 
turbulent velocity, and $V_A$ is the Alfv\'{e}n velocity. 
The form of $ I_A(k) $ reflects the scale-dependent anisotropy of Alfv\'{e}nic turbulence, with smaller turbulent eddies more elongated along the 
local magnetic field, i.e., $k_\perp \gg k_\|$. 
Because of the anisotropy, we approximately have 
\begin{equation}
    J_1(x) \approx \sqrt{\frac{2}{\pi x}}
\end{equation}
at a large $x$, as (Eqs. \eqref{eq: asux} and \eqref{eq: qltreg})
\begin{equation}
  k_{\perp,\text{res}} \gg k_{\|,\text{res}} \approx \frac{\Omega}{ v_\|} \sim \frac{\Omega} {v_\perp} = r_g^{-1},
\end{equation}
where $k_{\|,\text{res}} \approx \Omega/ v_\|$ and 
$k_{\perp,\text{res}} = L^\frac{1}{2} k_{\|,\text{res}}^\frac{3}{2}$ are the parallel and perpendicular resonant wavenumbers. 
Therefore the analytical reduction of Eq. \eqref{eq: oriduav} is 
\begin{align}
 &   D_{\mu\mu, \text{QLT}, A}  \nonumber\\
                                     \approx &  4\pi C_A C_\mu  \Omega^3 v_\perp^{-3} v_\|^{-1} 
                                      \int  dk_\perp     k_\perp^{-\frac{16}{3}}  
                                                    \exp{\Bigg(-L^\frac{1}{3}\frac{  \frac{\Omega}{v_\|}  }{k_\perp^\frac{2}{3}}\Bigg)}   \label{eq: intpeak} \\
       \approx &  \frac{2}{3} 8^\frac{13}{2} \exp (-8 ) \frac{\delta B_A^2}{B_0^2} \Big(\frac{v}{L \Omega}\Big)^\frac{3}{2} \frac{v}{L}  
    (1-\mu^2)^{-\frac{1}{2}}   \mu ^{\frac{11}{2}} .   \label{eq: duualffa}
\end{align}
The function in the integral in Eq. \eqref{eq: intpeak} peaks at 
\begin{equation}
   k_{\perp p} =  \Big( \frac{  L^\frac{1}{3} k_{\|,\text{res}} }{8}   \Big)^\frac{3}{2}   
   = 8^{-\frac{3}{2}} k_{\perp,\text{res}},
\end{equation}
which in fact is significantly smaller than $k_{\perp,\text{res}}$, but can still be much larger than $k_{\|, \text{res}}$ given $L \gg r_g$. 
With the disparity between $k_{\perp p}$ and $k_{\|, \text{res}}$, 
interactions with many uncorrelated eddies in the perpendicular direction over a gyro orbit  
are ineffective.

\begin{figure}[htbp]
\centering   
   \includegraphics[width=8.5cm]{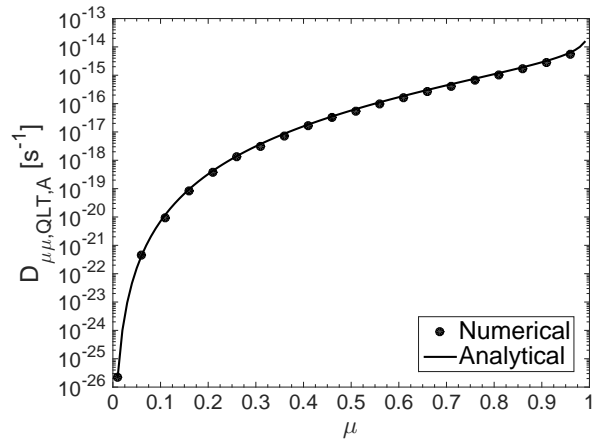}
\caption{ $D_{\mu\mu, \text{QLT}, A}$ as a function of $\mu$ for TeV CRs. 
{In the figure ``Numerical" indicates the numerical evaluation of Eq. \eqref{eq: oriduav}, and ``Analytical" indicates its analytical approximation 
given by Eq. \eqref{eq: duualffa}}.}
\label{fig: duuqltalf}
\end{figure}

As dictated by the turbulence anisotropy,
gyroresonant scattering by Alfv\'{e}n modes is inefficient. 
Compared with earlier studies, 
our result in Eq. \eqref{eq: duualffa} is different from that in 
\citet{Chan00}.
Based on the numerical simulations by 
\citet{CLV_incomp},
here we use the exponential form in the energy spectrum (Eq. \eqref{eq: enespa}), which was found to be more appropriate to describe the turbulence anisotropy
than the step function used in \citet{Chan00}.
Our formula is also simpler and more physically transparent than the one presented in  
\citet{YL02}.
Fig. \ref{fig: duuqltalf} illustrates $D_{\mu\mu, \text{QLT}, A}$ for TeV CRs, and here we adopt $L = 30$ pc and 
$\delta B_A = B_0 = 3 \mu G$. 
We note that for higher-energy CRs that interact with larger-scale turbulent eddies, 
since the turbulence anisotropy is weak, 
the assumption of a large $x$ at a large $\mu$ is invalid, and thus the 
approximate expression of $D_{\mu\mu,\text{QLT},A}$ in Eq. \eqref{eq: duualffa} is not applicable at a large $\mu$.

(2) Slow modes. 
In the case of the gyroresonant scattering by slow modes, there is 
\citep{Volk:1975}
\begin{equation}\label{eq: comgyd}
      D_{\mu\mu,\text{slow/fast}} = C_\mu \int d^3k \frac{k_\|^2}{k^2} [J_1^\prime (x)]^2 I(k) R(k)  .
\end{equation}
As mentioned earlier, slow modes are passively mixed by Alfv\'{e}n modes 
\citep{LG01,XJL19}
and thus follow the same scaling law as Alfv\'{e}n modes, with the energy spectrum 
\citep{CLV_incomp},
\begin{equation}\label{eq: slspe}
     I_s(k) = C_s  k_\perp^{-\frac{10}{3}} \exp{\Bigg(-L^\frac{1}{3}\frac{k_\|}{k_\perp^\frac{2}{3}}\Bigg)},
\end{equation} 
where 
\begin{equation}\label{eq: cs}
    C_s = \frac{1}{6 \pi} \delta B_s^2 L^{-\frac{1}{3}},
\end{equation}
and $\delta B_s$ is the rms strength of magnetic fluctuations of slow modes at $L$. 
To derive the approximate expression of $D_{\mu\mu,\text{QLT},s}$, we use 
\begin{equation}
     J_1^\prime (x) = \frac{1}{2} [J_0 (x) - J_2 (x) ]
      \approx  \sqrt{\frac{2}{\pi x}}
\end{equation}
at a large $x$ and assume $k^2 \sim k_\perp^2$ based on turbulence anisotropy, leading to 
\begin{align}
   & D_{\mu\mu, \text{QLT}, s} \nonumber\\
                                              \approx & 4\pi C_sC_\mu  \Omega^3 v_\perp^{-1} v_\|^{-3}  \int     dk_\perp   k_\perp^{-\frac{16}{3}} 
    \exp{\Bigg(-L^\frac{1}{3}\frac{\frac{ \Omega}{v_\|}}{k_\perp^\frac{2}{3}}\Bigg)} \\
    \approx & \frac{2}{3} 8^\frac{13}{2} \exp(-8) \frac{\delta B_s^2}{B_0^2} \Big(\frac{v}{L\Omega}\Big)^\frac{3}{2} \frac{v}{L} (1-\mu^2)^{\frac{1}{2}}  \mu^{\frac{7}{2}}   \label{eq: duusapsa} \\
    = &\frac{\delta B_s^2}{\delta B_A^2} \frac{1-\mu^2}{\mu^2} D_{\mu\mu, \text{QLT}, A}.
\end{align}
The above expression of $D_{\mu\mu, \text{QLT}, s}$ is similar to $D_{\mu\mu, \text{QLT}, A}$ in Eq. \eqref{eq: duualffa}, 
except for the different magnetic fluctuations of slow modes and the dependence on $\mu$. 
With comparable $\delta B_s$ and $\delta B_A$, 
$D_{\mu\mu, \text{QLT}, s}$ is larger than $D_{\mu\mu, \text{QLT}, A}$ at a small $\mu$, 
but smaller than $D_{\mu\mu, \text{QLT}, A}$ at a large $\mu$.
Our result in Eq. \eqref{eq: duusapsa} is also different from that in 
\citet{Chan00}, 
because of the different energy spectrum $I_s (k)$ (Eq. \eqref{eq: slspe}) used here.

\begin{figure}[htbp]
\centering   
   \includegraphics[width=8.5cm]{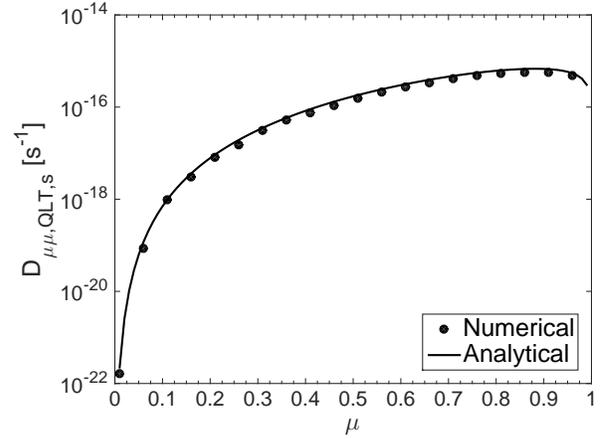}
\caption{ Same as Fig. \ref{fig: duuqltalf}, but for
$D_{\mu\mu, \text{QLT}, s}$. 
{``Numerical" indicates the numerical evaluation of Eq. \eqref{eq: comgyd}, and ``Analytical" indicates its analytical approximation 
given by Eq. \eqref{eq: duusapsa}}.}
\label{fig: duuqltslow}
\end{figure}

It shows that being subject to the same effect of turbulence anisotropy as Alfv\'{e}n modes,
gyroresonant scattering by slow modes is also inefficient. 
In Fig. \ref{fig: duuqltslow}, we present $D_{\mu\mu, \text{QLT}, s}$ for TeV CRs by using the same parameters as in 
Fig. \ref{fig: duuqltalf}
and assuming $\delta B_s = \delta B_A$. 
Similar to $D_{\mu\mu, \text{QLT}, A}$, the approximate expression of $D_{\mu\mu, \text{QLT}, s}$ in Eq. \eqref{eq: duusapsa}
does not apply to higher-energy CRs at a large $\mu$ due to the weak turbulence anisotropy on large scales.

\begin{figure}[htbp]
\centering   
   \includegraphics[width=8.5cm]{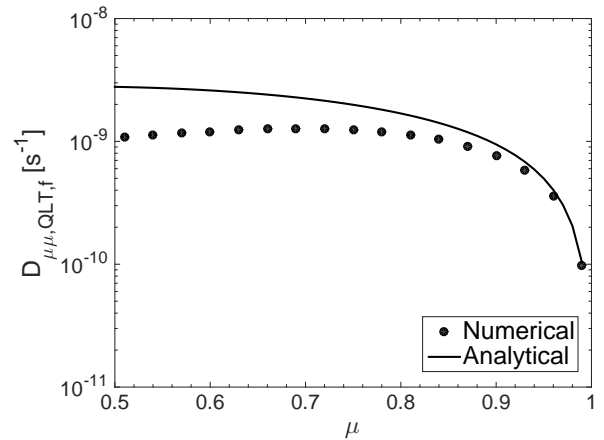}
\caption{ Same as Fig. \ref{fig: duuqltalf}, but for 
$D_{\mu\mu, \text{QLT}, f}$ at $\mu > 0.5$. 
{``Numerical" indicates the numerical evaluation of Eq. \eqref{eq: comgyd}, and ``Analytical" indicates its analytical approximation 
given by Eq. \eqref{eq: appfqltu}}.
}
\label{fig: duufastqlt}
\end{figure}

(3) Fast modes.
For gyroresonant scattering by fast modes, Eq. \eqref{eq: comgyd} also applies. 
Different from Alfv\'{e}n and slow modes, fast modes have isotropic scaling, and their energy spectrum is 
\citep{CL02_PRL}
\begin{equation}\label{eq: fsep}
     I_f(k) = C_f k^{-\frac{7}{2}},
\end{equation}
where 
\begin{equation}\label{eq: cf}
    C_f = \frac{1}{16 \pi} \delta B_f^2 L^{-\frac{1}{2}}, 
\end{equation}
and $\delta B_f$ is the rms strength of magnetic fluctuations of fast modes. 
By using the asymptotic expression 
\begin{equation}
     J_1^\prime (x) 
     \approx  \frac{1}{2} \Big(1 - \frac{x^2}{8} \Big)
\end{equation}
at a small $x$, we can obtain 
\citep{Xuc16,XLb18}
\begin{equation}\label{eq: appfqltu}
\begin{aligned}
      D_{\mu\mu,\text{QLT},f} 
         &  \approx \frac{ \pi}{56} \frac{ \delta B_f^2}{B_0^2}   \Big(\frac{v}{L\Omega}\Big)^\frac{1}{2}
                \Omega  (1 - \mu^2) \mu^{\frac{1}{2}}  .
\end{aligned}
\end{equation}
The comparison between the numerical and analytical results in Fig. \ref{fig: duufastqlt} shows that 
the above expression provides a better approximation for 
$D_{\mu\mu,\text{QLT},f}$ at a large $\mu$, where 
the assumption of a small $x$ is valid. 
In Fig. \ref{fig: duufastqlt}, we again adopt the same parameters as in Fig. \ref{fig: duuqltalf}
and assume $\delta B_f = \delta B_A$.

We see that due to the isotropic scaling, gyroresonant scattering by fast modes is efficient. 
In addition, 
$D_{\mu\mu,\text{QLT},f}$ 
decreases with decreasing $\Omega$ for CRs with higher energies,
while $D_{\mu\mu,\text{QLT},A}$ and $D_{\mu\mu,\text{QLT},s}$ increase with CR energy, 
since the anisotropy of Alfv\'{e}n and slow modes is weaker at larger scales.

\section{Trapping of CRs and its effect on diffusion}

\subsection{Trapping by magnetic compressions}

Compressive magnetic fluctuations induced by fast and slow modes in compressible MHD turbulence 
and pseudo-Alfv\'{e}n modes, which are the incompressible limit of slow modes, 
act as magnetic mirrors. 
The large-scale magnetic compressions with the variation wavenumber $k < r_g^{-1}$
give rise to the adiabatic trapping of CRs. 
As the magnetic moment can be treated as an adiabatic invariant, we have 
\begin{equation}
     \frac{v_{\perp}^2}{B_0}  =  \frac{v^2}{B_0+ b_k},
\end{equation}
where $b_k$ is the compressive magnetic fluctuation at $k$.
It follows that the angular size of the loss cone $\theta_{lc}$ satisfies  
\begin{equation}
      \mu_{lc}^2 =  \cos^2 \theta_{lc} 
       = \frac{ b_k }{B_0 + b_k} \approx \frac{ b_k }{B_0 }
\end{equation}
when $b_k \ll B_0$. 
The particles with $\mu < \mu_{lc}$ are subject to trapping.

The mirror force exerted by fast modes on a trapped particle is 
\begin{equation}
    \frac{d (p \mu)}{ dt} =- \frac{p_\perp v_\perp}{ 2 B_0 } b_{fk} k    , 
\end{equation}
where $p$ is the particle momentum, and $b_{fk}$ is $b_k$ of fast modes. 
Then the rate of change in $\mu$ due to trapping, which we term as the trapping rate, has the form
\citep{CesK73}
\begin{equation}
   \Gamma_{t,f} = \Big |\frac{1}{\mu} \frac{d\mu}{dt} \Big| =  \frac{v}{2B_0} \frac{1-\mu^2}{\mu}  b_{fk} k.
\end{equation}
Among the magnetic mirrors at different wavenumbers, the ones that 
are most effective in reflecting the CR particle at a given $\mu$ have 
\citep{CesK73}
\begin{equation}
    b_{fk} = B_0 \mu^2, 
\end{equation}
for which the inverse of $\Gamma_{t,f}$ is just the time for a particle to bounce between reflection points.
By further using the scaling of fast modes 
\citep{CL02_PRL}
\begin{equation}
    b_{fk} = \delta B_f (kL)^{-\frac{1}{4}}, 
\end{equation}
we finally reach $\Gamma_{t,f}$ as a function of $\mu$,
\begin{equation}\label{eq: gtraft}
   \Gamma_{t,f}  =  \frac{v}{2L} \frac{\delta B_f^4}{B_0^{4} } \frac{1-\mu^2}{\mu^7}   .
\end{equation}
It rapidly decreases with increasing $\mu$, 
as the mirror reflection is slower at a smaller $k$.

{In the above expression of $\Gamma_{t,f}$, the minimum $\mu$ for the adiabatic trapping of CRs by fast modes should satisfy 
\begin{equation}
     \mu_\text{min,f}^2 = \frac{b_{fk} (r_g)}{B_0}  = \frac{\delta B_f}{B_0} \Big( \frac{r_g}{L} \Big)^{\frac{1}{4}},
\end{equation}
where $b_{fk} (r_g)$ is the magnetic fluctuation of fast modes at $r_g$. 
Since the magnetic compressions at $k > r_g^{-1}$ are incapable of trapping, 
$ \Gamma_{t,f}$ at $\mu < \mu_\text{min,f}$ is in fact given by 
\begin{equation}\label{eq: trfasmu}
\begin{aligned}
     \Gamma_{t,f} (\mu < \mu_\text{min,f}) & =   \frac{v}{2B_0} \frac{1-\mu^2}{\mu}  b_{fk}(r_g) r_g^{-1} \\
                                                                  & =  \frac{v}{2 r_g } \frac{ \delta B_f}{B_0} \Big( \frac{r_g}{L} \Big)^{\frac{1}{4}}  \frac{1-\mu^2}{\mu} .
\end{aligned}
\end{equation}
In addition, as the compressive fluctuations move with a phase speed $V_\text{ph,f}$, 
when the parallel particle speed $v_\|$ becomes smaller than $V_\text{ph,f}$ with $\mu \lesssim V_\text{ph,f}/ v$, 
the above formulae of $\Gamma_{t,f}$ in the magnetostatic limit are inapplicable. 
However,  given $V_\text{ph,f} \ll v$ for non-relativistic MHD turbulence, the above formulae of $\Gamma_{t,f}$ can be safely used 
except for $\mu \rightarrow 0$.}

In the case of slow modes or pseudo-Alfv\'{e}n modes, 
the motion of a particle along the magnetic field is described by 
\begin{equation}
    \frac{d (p \mu)}{ dt} = -\frac{p_\perp v_\perp}{ 2 B_0 } b_{sk} k_\|,
\end{equation}
where $b_{sk}$ is $b_k$ of slow modes or pseudo-Alfv\'{e}n modes.
Thus the trapping rate is 
\begin{equation}
    \Gamma_{t,s} =  \Big| \frac{1}{\mu} \frac{d\mu}{dt} \Big|=  \frac{v}{2B_0} \frac{1-\mu^2}{\mu}  b_{sk} k_\|.
\end{equation}
Similar to the case of fast modes, 
under the consideration of both 
\begin{equation}
   b_{sk} = B_0 \mu^2
\end{equation}
and the scaling of slow modes/pseudo-Alfv\'{e}n modes
\citep{CLV_incomp}
\begin{equation}\label{eq: anislber}
    b_{sk} = \delta B_s (k_\perp L)^{-\frac{1}{3}}
               = \delta B_s (k_\| L)^{-\frac{1}{2}},
\end{equation}
we find the trapping rate of slow/pseudo-Alfv\'{e}n modes as 
\begin{equation}\label{eq: spsegtr}
    \Gamma_{t,s} 
    =  \frac{v}{2 L } \frac{\delta B_s^2}{B_0^2} \frac{1-\mu^2}{\mu^3}     .
\end{equation}
{The above expression is valid for $\mu > \mu_\text{min,s}$, where 
\begin{equation}
   \mu_\text{min,s}^2 =  \frac{b_{sk} (r_g)}{B_0} = \frac{\delta B_s}{B_0} \Big(\frac{r_g}{ L}\Big)^{\frac{1}{2}},
\end{equation}
and $b_{sk} (r_g)$ is the magnetic fluctuation of slow modes at $k_\| = 1/ r_g$. 
$ \Gamma_{t,s} $ at $\mu < \mu_\text{min,s}$ is given by 
\begin{equation}\label{eq: trfasmusl}
\begin{aligned}
    \Gamma_{t,s} (\mu < \mu_\text{min,s}) & = \frac{v}{2B_0} \frac{1-\mu^2}{\mu}  b_{sk} (r_g) r_g^{-1} \\
                                                                  & = \frac{v}{2 r_g }  \frac{\delta B_s}{B_0} \Big(\frac{r_g}{ L}\Big)^{\frac{1}{2}}  \frac{1-\mu^2}{\mu}.
\end{aligned}     
\end{equation}
Similar to the case of fast modes, 
the above formulae do not apply to $\mu \lesssim V_\text{ph,s}/ v$,
where $V_\text{ph,s} \ll v$ is the phase speed of slow modes. }

By comparing $\Gamma_{t,s}$ in Eq. \eqref{eq: spsegtr} with $\Gamma_{t,f}$ in Eq. \eqref{eq: gtraft}, we see that 
\begin{equation}
     \frac{\Gamma_{t,f}}{\Gamma_{t,s}} = \frac{\delta B_f^4}{B_0^2 \delta B_s^2} \mu^{-4}.
\end{equation}
With the anisotropic energy distribution, 
$b_{sk}$ decreases more rapidly in the direction parallel to the magnetic field 
(Eq. \eqref{eq: anislber}, see also 
\citealt{Bere15}).
Because of the different scalings of fast and slow modes, 
$\Gamma_{t,f}$ is much larger than $\Gamma_{t,s}$
when $\delta B_f$ and $\delta B_s$ are comparable to $B_0$.

\subsection{Diffusion in compressible and incompressible MHD turbulence}

\begin{figure}[htbp]
\centering   
   \includegraphics[width=9cm]{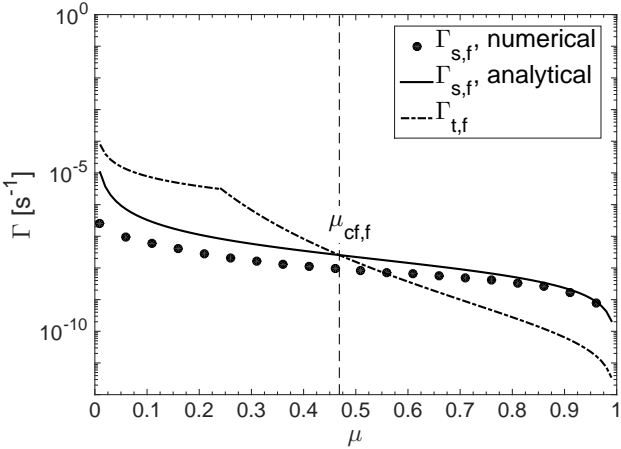}
\caption{ Comparison between $\Gamma_{t,f}$ and $\Gamma_{s,f}$ in compressible MHD turbulence for TeV CRs.
{The analytical approximations are given by Eqs. \eqref{eq: gtraft}, \eqref{eq: trfasmu}, and \eqref{eq: scraft}. 
The numerical calculation of Eq. \eqref{eq: comgyd} is used for the numerical evaluation of $\Gamma_{s,f}$.}
The vertical dashed line indicates $\mu_\text{cf,,f}$, corresponding to Eq. \eqref{eq: faucf}.  }
\label{fig: mucff}
\end{figure}

In the presence of trapping, 
the diffusion of CRs only occurs when the 
pitch-angle scattering can overcome the magnetic trapping. 
The rate of change in $\mu$ due to scattering, i.e., scattering rate $\Gamma_s$, is related to $D_{\mu\mu}$
\citep{Jokipii1966},
\begin{equation}\label{eq: gensr}
    \Gamma_s = \frac{1}{\mu^2} \frac{\langle (\Delta \mu)^2 \rangle}{\Delta t} 
                       =  \frac{2 D_{\mu\mu}}{\mu^2} .
\end{equation}
In compressible MHD turbulence, fast modes dominate the pitch-angle scattering
(Section \ref{sec: pasca}).
The scattering rate of fast modes is 
\begin{equation}\label{eq: scraft}
    \Gamma_{s,f} = \frac{2 D_{\mu\mu,\text{QLT},f}}{\mu^2} \approx
          \frac{ \pi}{28} \frac{ \delta B_f^2}{B_0^2}   \Big(\frac{v}{L\Omega}\Big)^\frac{1}{2}
                \Omega  (1 - \mu^2) \mu^{-\frac{3}{2}}  .
\end{equation}
As fast modes dominate both trapping and scattering in compressible MHD turbulence, the comparison between 
$\Gamma_{t,f}$ and $\Gamma_{s,f}$ determines the range of $\mu$ where the CRs mainly contribute to the spatial diffusion. 
The cutoff pitch-angle cosine is defined at the balance between $\Gamma_{t,f}$ and $\Gamma_{s,f}$
\citep{CesK73}. 
It has the expression (Eqs. \eqref{eq: gtraft} and \eqref{eq: scraft})
\begin{equation}\label{eq: faucf}
    \mu_\text{cf,f} \approx  \bigg[ \frac{14}{\pi} \frac{ \delta B_f^2}{B_0^2} \Big(\frac{v}{ L \Omega}\Big)^\frac{1}{2} \bigg]^\frac{2}{11}.
\end{equation}
Fig. \ref{fig: mucff} displays both $\Gamma_{t,f}$ and $\Gamma_{s,f}$ for TeV CRs and their intersection at $\mu_\text{cf,f}$. 
The analytical approximation of $\Gamma_{s,f}$ agrees well with its numerical value toward a larger $\mu$. 
At $\mu < \mu_\text{cf,f}$, CRs are mainly reflected back and forth between mirror points due to the dominant trapping effect. 
At $\mu> \mu_\text{cf,f}$, scattering becomes more important than trapping and enables diffusion of CRs. 

\begin{figure}[htbp]
\centering   
   \includegraphics[width=9cm]{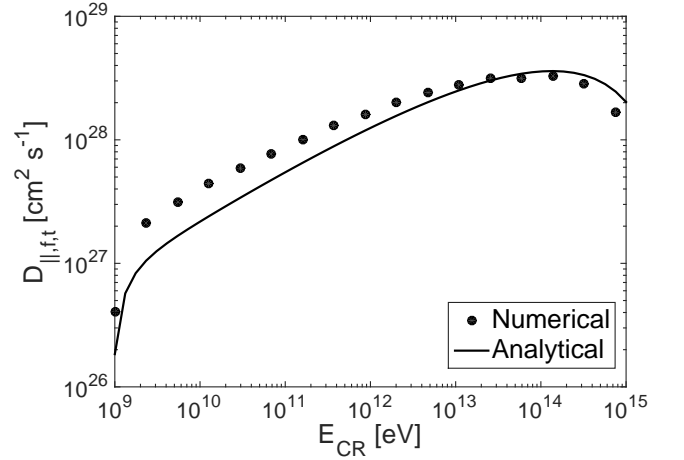}
\caption{ $D_{\|,f,t}$ as a function of $E_\text{CR}$.  
The analytical approximation is given by Eqs. \eqref{eq: fapasdt} and \eqref{eq: faucf}. }
\label{fig: fastdp}
\end{figure}

As a result, the parallel spatial diffusion coefficient of CRs due to the gyroresonant scattering by fast modes 
should be calculated as 
\begin{align}
    D_{\|,f,t} &= \frac{1}{3} v \lambda_\| 
    =  \frac{v^2}{4 } \int_{\mu_\text{cf,f}}^{1} d\mu \frac{(1-\mu^2)^2}{D_{\mu\mu,\text{QLT},f}}  \label{eq: fapasnumdt} \\
    &\approx \frac{14}{\pi} \frac{B_0^2}{ \delta B_f^2} \Big(\frac{v}{L\Omega}\Big)^{-\frac{1}{2}} \frac{v^2}{\Omega}
     \int_{\mu_\text{cf,f}}^{1} d\mu \frac{1-\mu^2}{  
                   \mu^{\frac{1}{2}} }  \nonumber\\
    &= \frac{28}{5\pi} \frac{B_0^2}{ \delta B_f^2} \Big(\frac{v}{L\Omega}\Big)^{-\frac{1}{2}} \frac{v^2}{\Omega}
      \big[4- \sqrt{\mu_\text{cf,f}} (5-\mu_\text{cf,f}^2 )\big] , \label{eq: fapasdt}
\end{align}
{where Eq. \eqref{eq: appfqltu} is used in deriving Eq. \eqref{eq: fapasdt}.}
Instead of an integration over the entire range of pitch angles, 
here the lower limit of the integral 
is determined by $\mu_\text{cf,f}$. 
As a function of $\mu_\text{cf,f}$, 
$D_{\|,f,t}$ for a more energetic CR is more significantly affected by trapping 
as $\mu_\text{cf,f}$ increases with the CR energy (Eq. \eqref{eq: faucf}).
We see from Fig. \ref{fig: fastdp} that under the effect of trapping, 
$D_{\|,f,t}$ has a weaker dependence on CR energy $E_\text{CR}$ toward higher energies. 
{The discrepancy between the analytical approximation (Eq. \eqref{eq: fapasdt}) and the numerical evaluation of Eq.\eqref{eq: fapasnumdt}
mainly comes from the overestimate of $D_{\mu\mu,\text{QLT},f}$ by Eq. \eqref{eq: appfqltu}. 
The drop of $D_{\|,f,t}$ near $1$ GeV is due to the significant drop of $v$ at $1$ GeV. }

\begin{figure*}[htbp]
\centering   
\subfigure[]{
   \includegraphics[width=8.5cm]{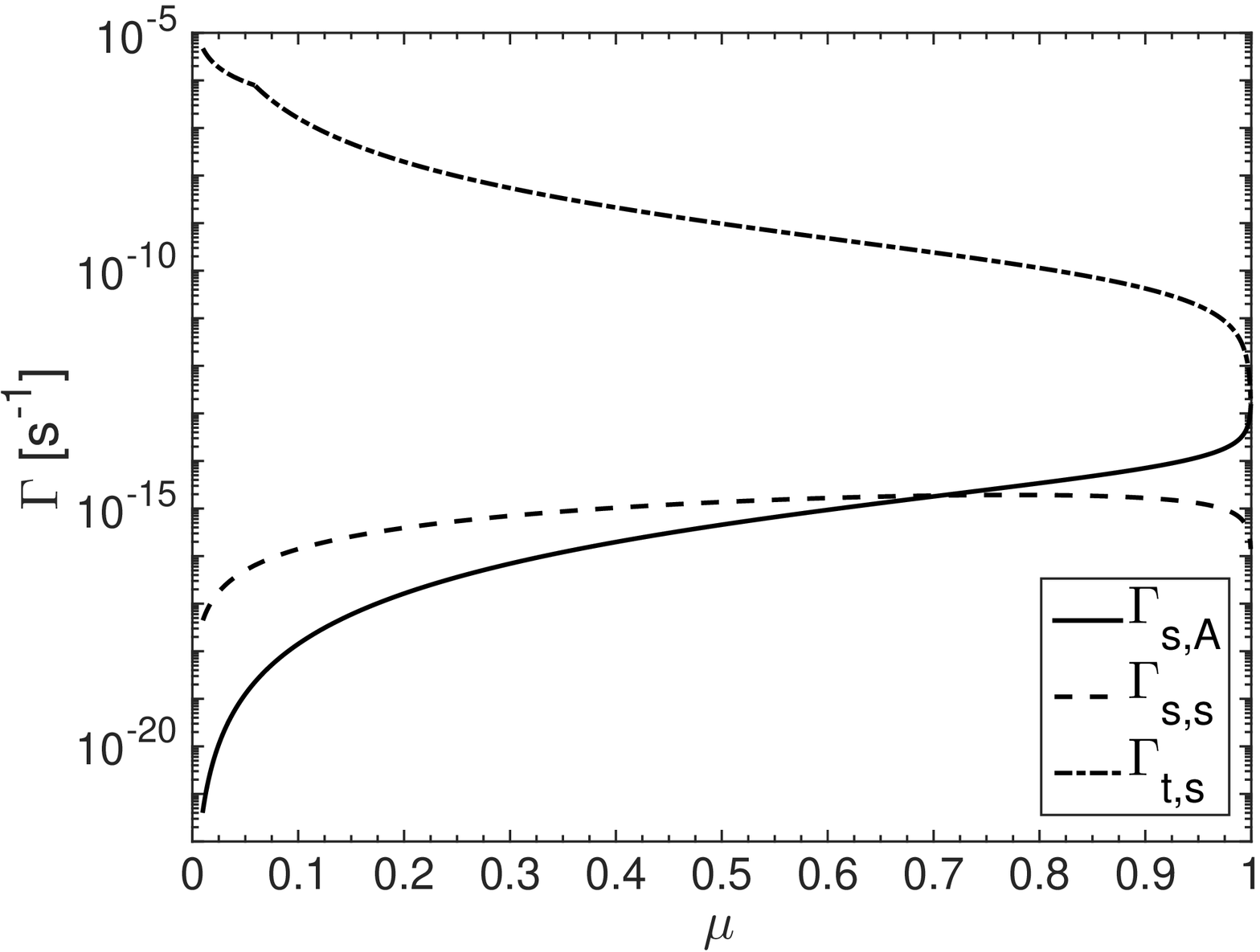}\label{fig: ucfas}}
\subfigure[]{
   \includegraphics[width=8.5cm]{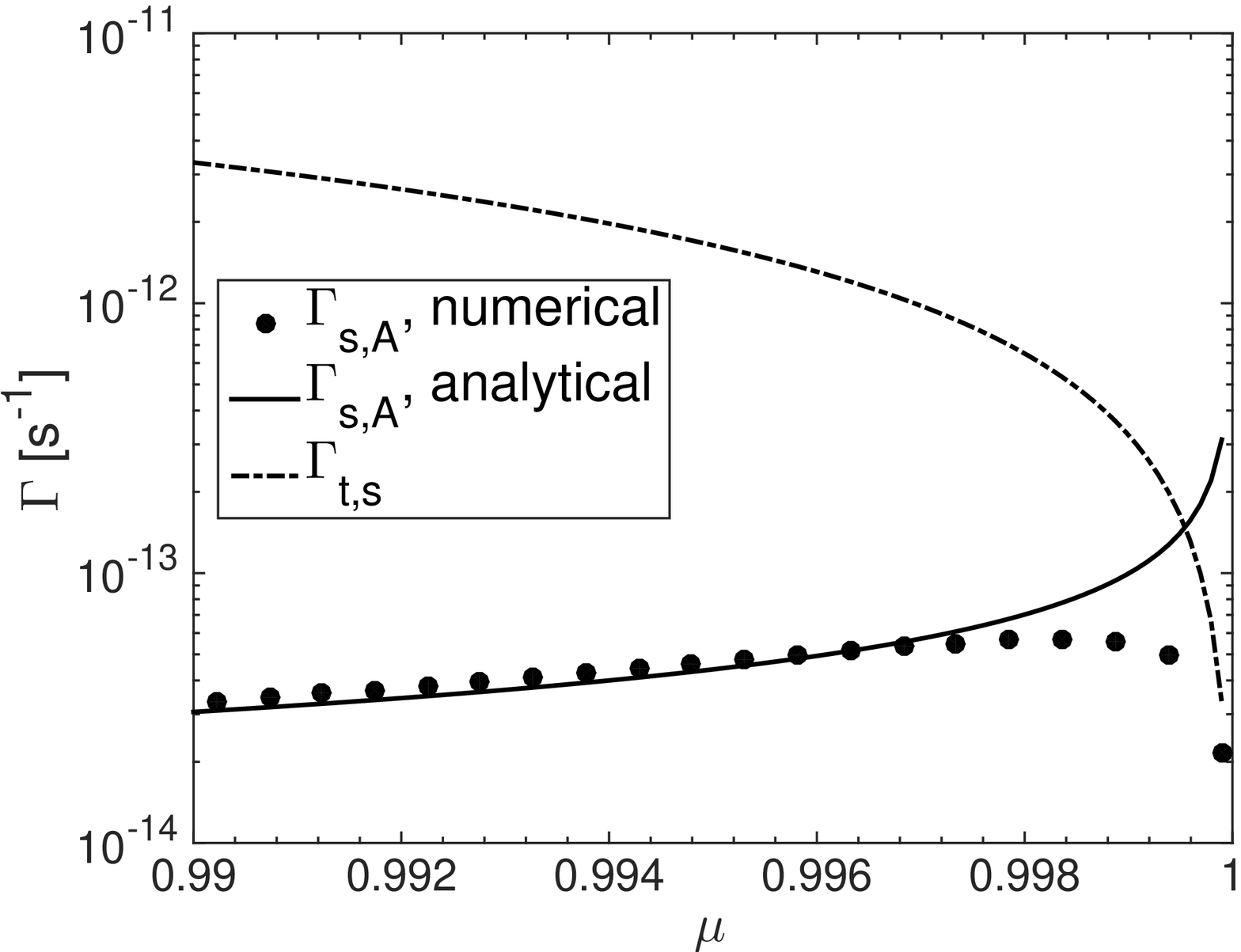}\label{fig: ucfasz}}
\caption{ (a) $\Gamma_{t,s}$ in comparison with $\Gamma_{s,A}$ and $\Gamma_{s,s}$ in incompressible MHD turbulence for TeV CRs, 
with the {analytical expressions given by Eqs. \eqref{eq: spsegtr}, \eqref{eq: trfasmusl}, \eqref{eq: gascalf}, and \eqref{eq: gascrslw}.}
(b) Zoom of (a) near $\mu=1$. 
{The numerical evaluation of $\Gamma_{s,A}$ is obtained by using the numerical calculation of $D_{\mu\mu,\text{QLT},A}$
in Eq. \eqref{eq: oriduav}. }} 
\label{fig: slmucf}
\end{figure*}

In incompressible MHD turbulence, the pseudo-Alfv\'{e}n modes give rise to trapping. 
On the other hand, the scattering rate of Alfv\'{e}n modes is (Eqs. \eqref{eq: duualffa} and \eqref{eq: gensr})
\begin{equation}\label{eq: gascalf}
\begin{aligned}
     \Gamma_{s,A}&= \frac{2 D_{\mu\mu,\text{QLT},A}}{\mu^2}  \\
    & \approx
     \frac{4}{3} 8^\frac{13}{2} \exp (-8 ) \frac{\delta B_A^2}{B_0^2} \Big(\frac{v}{L \Omega}\Big)^\frac{3}{2} \frac{v}{L}  
    (1-\mu^2)^{-\frac{1}{2}}   \mu ^{\frac{7}{2}} .
\end{aligned}
\end{equation}
The scattering rate of slow modes is (Eqs. \eqref{eq: duusapsa} and \eqref{eq: gensr})
\begin{equation}\label{eq: gascrslw}
\begin{aligned}
     \Gamma_{s,s}&= \frac{2 D_{\mu\mu,\text{QLT},s}}{\mu^2}  \\
     & \approx 
      \frac{4}{3} 8^\frac{13}{2} \exp(-8) \frac{\delta B_s^2}{B_0^2} \Big(\frac{v}{L\Omega}\Big)^\frac{3}{2} \frac{v}{L} (1-\mu^2)^{\frac{1}{2}}  \mu^{\frac{3}{2}}  .
\end{aligned}
\end{equation}
Given $\delta B_s$ comparable to $\delta B_A$, $\Gamma_{s,s}$ is larger than $\Gamma_{s,A}$ at a small $\mu$, 
but smaller than $\Gamma_{s,A}$ at a larger $\mu$.
Fig. \ref{fig: slmucf} presents $\Gamma_{t,s}$ in comparison with $\Gamma_{s,A}$ and $\Gamma_{s,s}$ for TeV CRs. 
It show that 
over the entire range of pitch angles, trapping dominates over scattering. 
Consequently, TeV CRs are trapped by pseudo-Alfv\'{e}n modes and 
prevented from participating in diffusion.
In Appendix \ref{app: slttd}, we consider the 
resonance-broadened transit time damping (TTD) with pseudo-Alfv\'{e}n modes 
\citep{XLb18}
as a mechanism to enhance the pitch-angle scattering of CRs. 
It turns out that the TTD with resonance broadening can still be insufficient to overcome the trapping in incompressible MHD turbulence, 
{depending on the turbulence parameters.}

\begin{figure}[htbp]
\centering   
   \includegraphics[width=9cm]{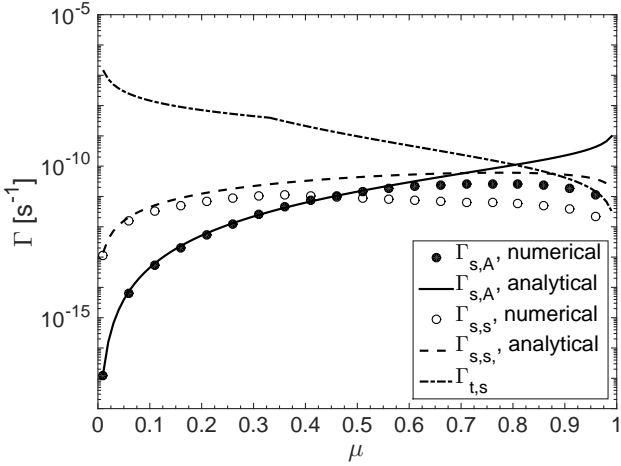}
\caption{ Same as Fig. \ref{fig: ucfas} but for PeV CRs. 
{The numerical evaluations of $\Gamma_{s,A}$ and $\Gamma_{s,s}$ are 
obtained by using the numerical calculations of $D_{\mu\mu,\text{QLT},A}$
in Eq. \eqref{eq: oriduav} and $D_{\mu\mu,\text{QLT},s}$ in Eq. \eqref{eq: comgyd}, respectively. }}
\label{fig: pevcrinc}
\end{figure}

For higher-energy CRs, although the gyroresonant scattering by Alfv\'{e}n and pseudo-Alfv\'{e}n modes becomes more efficient (see Section \ref{sec: pasca}), 
trapping still dominates over scattering for the entire range of $\mu$. 
Fig. \ref{fig: pevcrinc} presents $\Gamma_{t,s}$ in comparison with $\Gamma_{s,A}$ and $\Gamma_{s,s}$ for PeV CRs.
We note that the analytical approximations Eqs \eqref{eq: gascalf} and \eqref{eq: gascrslw} are invalid for high-energy CRs at a large $\mu$ (see Section \ref{sec: pasca}).
It suggests that the diffusion of CRs in incompressible MHD turbulence
is hindered by trapping. 
Even with weak scattering, 
the motion of CRs is not ballistic in incompressible MHD turbulence.

In the above calculations, we assume that Alfv\'{e}n, slow, and fast modes have comparable magnetic fluctuations at $L$. 
In realistic astrophysical conditions, the energy fractions of different modes depend on the turbulence driving and the 
conversion from compressive to solenoidal motions along the cascade 
\citep{Pad16}. 
Besides, in weakly ionized interstellar phases, 
fast modes are subject to severe ion-neutral collisional damping 
\citep{Xuc16}. 
{All these effects should be taken into account to realistically model the 
trapping and diffusion of CRs in the multi-phase interstellar medium. }

\section{Discussion}

Recent advances in theoretical understanding of MHD turbulence have brought 
substantial changes of the paradigm of CR propagation. 
{Here we focus on relatively high-energy CRs that mainly interact with background MHD turbulence 
instead of the CR-induced streaming instability 
\citep{Lerche,Kulsrud_Pearce}.}
Our study suggests that a significant change is expected when the trapping of CRs is taken into account. 
In incompressible MHD turbulence, 
instead of the ballistic motion of CRs along magnetic field lines due to the weak scattering  
\citep{Chan00,YL02},
CRs remain trapped. 
For the compressible MHD turbulence in the ISM, 
earlier studies showed that the ballistic motion of CRs can be prevented due to the efficient scattering by fast modes
\citep{YL02}. 
In fact, besides scattering, 
fast modes also play a dominant role in trapping CRs. 
The propagation of CRs is controlled by 
the interplay between scattering and trapping by fast modes.

{In this work we only consider non-relativistic MHD turbulence. 
The similarity between non-relativistic and relativistic MHD turbulence has been found 
\citep{Thom98,Cho05,Zra12,Choim14,Tak15}.
A detailed study on the stochastic particle acceleration in relativistic MHD turbulence was recently 
carried out by 
\citet{Dem19}, 
which can be generally applied to high-energy astrophysical environments involving relativistic MHD turbulence 
(see, e.g., \citealt{XZg17,Xuyz18}).
The current study should be extended to the regime of relativistic MHD turbulence. }

Molecular-cloud magnetic mirrors were discussed in, 
e.g., \citet{Chanmc00},
for trapping and confining Galactic CRs. 
With the 
weak correspondence between magnetic fields and gas densities in most volumes of molecular clouds due to reconnection diffusion
\citep{Crut10,LEC12},
the existence of molecular-cloud mirrors is questionable. 
Here we consider the magnetic mirrors naturally arising in MHD turbulence, 
with different sizes and magnetic fluctuations resulting from the cascade of MHD turbulence. 
They are ubiquitous in the turbulent and magnetized ISM and do not depend on molecular cloud structures. 
In addition, different from the linear description of MHD waves adopted in, e.g., 
\citet{CesK73}, 
we use a realistic model of MHD turbulence. 
It turns out that both scattering and trapping of CRs strongly depend on the properties and scalings of MHD turbulence.
For instance, the anisotropy of incompressible MHD turbulence leads to inefficient scattering and 
significant trapping.

The trapping of CRs in MHD turbulence 
has important astrophysical implications. 
For instance, the second-order Fermi acceleration in the presence of trapping can be more efficient 
than the case with only pitch-angle scattering. 
The reflection of particles within magnetic traps can also give rise to 
more efficient shock acceleration than the diffusive shock acceleration with random scattering. 
Besides, the trapping of CRs {may also} significantly affect, e.g., the confinement of CRs in galaxies, 
the coupling of CRs to gas and galactic wind driving.
These implications will be addressed in future work.

\section{Summary}

In this work, we have applied the modern theories of MHD turbulence to studying the trapping of CRs in MHD turbulence. 
Our main results are as follows. 

1. The pitch-angle diffusion coefficients for gyroresonant scattering by Alfv\'{e}n and slow modes have similar formulae, 
which are inefficient for CRs with $r_g \ll L$ and increase with the energy of CRs due to the scale-dependent turbulence anisotropy. 
The more efficient gyroresonant scattering by isotropic fast modes has the pitch-angle diffusion coefficient decrease with the energy of CRs. 

2. The trapping of CRs by slow modes (or pseudo-Alfv\'{e}n modes in the incompressible limit) has a lower rate than that by fast modes 
due to the anisotropy of slow modes.

3. In compressible MHD turbulence, in the presence of trapping by fast modes, the gyroresonant scattering of CRs by fast modes 
can only occur within a limited range of pitch angles, resulting in the suppression of parallel diffusion of CRs. 
The dependence of the parallel diffusion coefficient on the CR energy becomes weaker toward higher energies.

4. In incompressible MHD turbulence, the trapping by pseudo-Alfv\'{e}n modes dominates over the scattering 
over the entire range of pitch angles, which inhibits the parallel diffusion of CRs. 
\\
\\
\\
S.X. acknowledges the support for Program number HST-HF2-51400.001-A provided by NASA through a grant from the Space Telescope Science Institute, which is operated by the Association of Universities for Research in Astronomy, Incorporated, under NASA contract NAS5-26555.
A.L. acknowledges the support from grant
NASA      
TCAN 144AAG1967.

\appendix 

\section{Resonance-broadened TTD with pseudo-Alfv\'{e}n modes in incompressible MHD turbulence }
\label{app: slttd}

The pitch-angle diffusion coefficient of pseudo-Alfv\'{e}n modes for TTD is 
\citep{Volk:1975},
\begin{equation}\label{eq:gel}
     D_{\mu\mu,\text{TTD},s} = C_\mu \int d^3k \frac{k_\|^2}{k^2} [J_0^\prime (x)]^2 I(k) R(k)  .
\end{equation}
Here we adopt a broadened resonance function
\citep{YL08,XLb18}, 
\begin{equation}\label{eq: vbmf}
    R_{B} = \frac{\sqrt{2\pi}}{2  \Delta v_\| k_\|}
    \exp{\Bigg[-\frac{(\omega_k - v_\| k_\|)^2 }{2 ( \Delta v_\| k_\|)^2}\Bigg]} ,
\end{equation}
where 
\begin{equation}\label{eq: volk}
     \Delta v_\|  \approx v_\perp \Bigg(\frac{\langle\delta B_\|^2\rangle}{B_0^2}\Bigg)^\frac{1}{4} ,
\end{equation}
is the variation in $v_\|$ induced by the parallel magnetic fluctuation $\delta B_\|$
\citep{Volk:1975}. 
The existence of $ \Delta v_\| $ causes the resonance broadening. 

\begin{figure*}[htbp]
\centering   
\subfigure[]{
   \includegraphics[width=8.5cm]{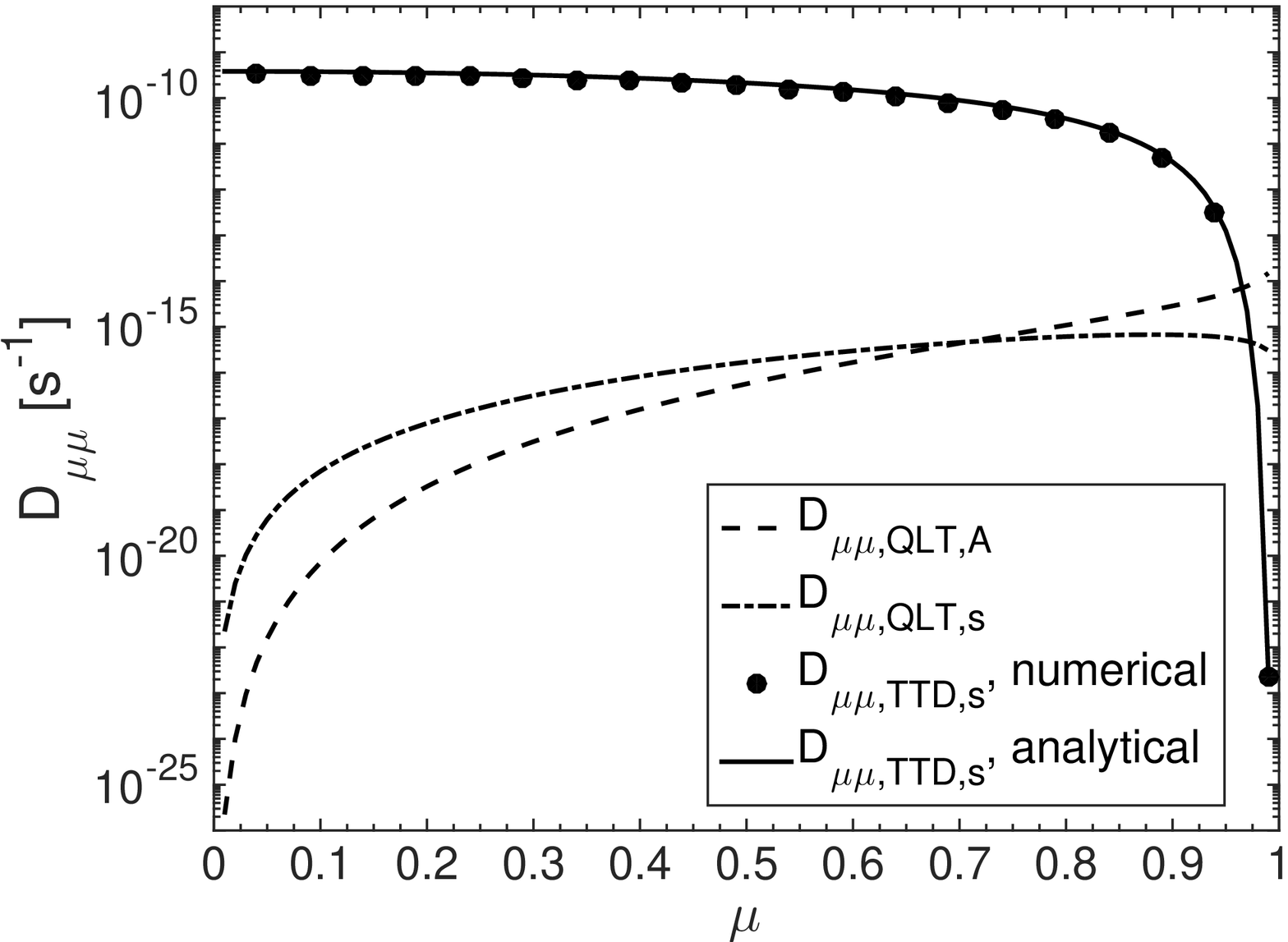}\label{fig: sttd}}
\subfigure[]{
   \includegraphics[width=8.5cm]{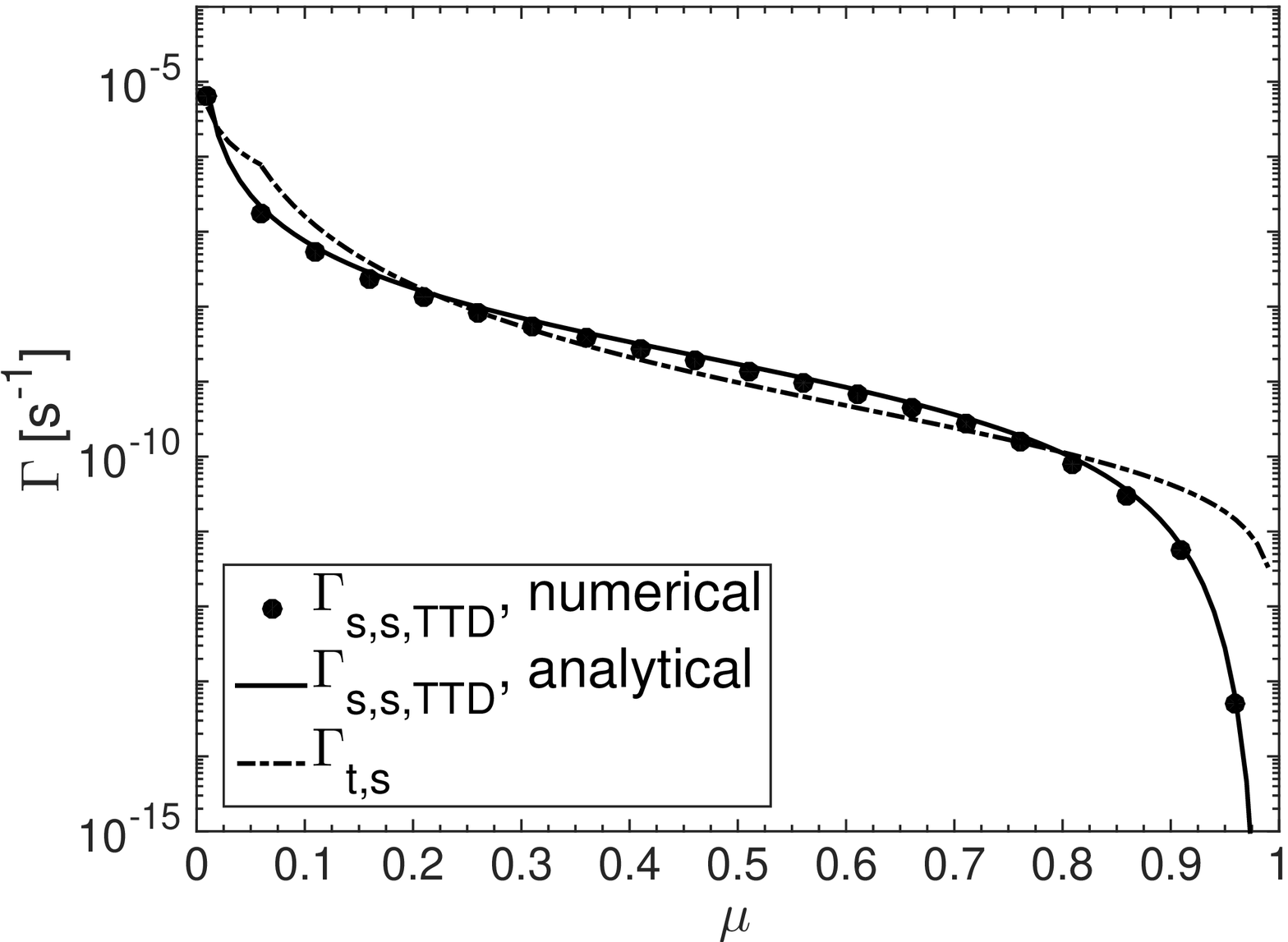}\label{fig: sttdtp}}
\caption{ (a) The pitch-angle diffusion coefficients for gyroresonance, $D_{\mu\mu,\text{QLT},A}$ (Alfv\'{e}n modes), $D_{\mu\mu,\text{QLT},s}$ (pseudo-Alfv\'{e}n modes),
and for resonance-broadened TTD, $D_{\mu\mu,\text{TTD},s}$ (pseudo-Alfv\'{e}n modes), for TeV CRs. 
{The numerical evaluation of $D_{\mu\mu,\text{TTD},s}$ is derived from the numerical calculation of Eq. \eqref{eq:gel}. 
Its} analytical approximation is taken from Eq. \eqref{eq: sbfurs}.
(b) The corresponding $\Gamma_{s,s,\text{TTD}}$ with the analytical approximation given by Eq. \eqref{eq: ttsgssta} 
in comparison with $\Gamma_{t,s}$ {(Eqs. \eqref{eq: spsegtr} and \eqref{eq: trfasmusl}).}}
\label{fig: slottd}
\end{figure*}

The approximate expression of $D_{\mu\mu,\text{TTD},s}$ for CRs is 
\citep{XLb18}
\begin{equation}\label{eq: sbfurs}
     D_{\mu\mu,\text{TTD},s} \approx  \frac{\sqrt{2}}{4} \pi^\frac{3}{2} \frac{C_s}{B_0^2} \Bigg(\frac{\langle\delta B_\|^2\rangle}{B_0^2}\Bigg)^{-\frac{1}{4}}
     L^{-\frac{2}{3}} \ln \Big(\frac{L}{l_{\perp,\text{min}}}\Big) v  
     (1-\mu^2)^\frac{3}{2} \exp \Bigg[-\frac{v_\|^2}{2\Delta v_\|^2}\Bigg] ,
\end{equation}
where $l_{\perp,\text{min}}$ is determined by the larger value between $r_g$ and the dissipation scale of magnetic fluctuations. 
Fig. \ref{fig: sttd} illustrates 
$D_{\mu\mu,\text{TTD},s}$ in comparison with the diffusion coefficients for gyroresonant scattering, i.e.,  
$D_{\mu\mu,\text{QLT},A}$ and $D_{\mu\mu,\text{QLT},s}$ presented in Figs. \ref{fig: duuqltalf} and \ref{fig: duuqltslow}, 
where $\delta B_\| = \delta B_s$ is used, 
and other parameters are the same as in Section \ref{sec: pasca}.
It shows that TTD dominates the pitch-angle scattering 
except for large $\mu$. 
The scattering rate corresponding to $D_{\mu\mu,\text{TTD},s}$ is 
\begin{equation}\label{eq: ttsgssta}
     \Gamma_{s,s,\text{TTD}} = \frac{2 D_{\mu\mu,\text{TTD},s} }{\mu^2} 
     \approx  \frac{\sqrt{2}}{2} \pi^\frac{3}{2} \frac{C_s}{B_0^2} \Bigg(\frac{\langle\delta B_\|^2\rangle}{B_0^2}\Bigg)^{-\frac{1}{4}}
     L^{-\frac{2}{3}} \ln \Big(\frac{L}{l_{\perp,\text{min}}}\Big) v  
     (1-\mu^2)^\frac{3}{2} \exp \Bigg[-\frac{v_\|^2}{2\Delta v_\|^2}\Bigg]  \mu^{-2}.
\end{equation}
As shown in Fig. \ref{fig: sttdtp}, 
$\Gamma_{s,s,\text{TTD}}$ is comparable to $\Gamma_{t,s}$ except for large $\mu$, where 
it is much smaller than $\Gamma_{t,s}$.
{We see that given the parameters used here,} although TTD with broadened resonance leads to more efficient scattering than gyroresonance, 
it is still insufficient to significantly untrap CRs.

\bibliographystyle{apj.bst}
\bibliography{xu}

\end{document}